# *Q*-dependence of the spin fluctuations in the intermediate valence compound CePd$_3$

V R Fanelli[1,2], J M Lawrence[1,2], E A Goremychkin[3], R Osborn[4], E D Bauer[1], K J McClellan[1], J D Thompson[1], C H Booth[5], A D Christianson[6] and P S Riseborough[7]

[1] Los Alamos National Laboratory, Los Alamos, NM 87545, USA
[2] University of California, Irvine, CA 92697, USA
[3] University of Southampton, Southampton S017 1BJ, United Kingdom
[4] Argonne National Laboratory, Argonne, IL 60439, USA
[5] Lawrence Berkeley National Laboratory, Berkeley, CA 94720, USA
[6] Oak Ridge National Laboratory, Oak Ridge, TN 37831, USA
[7] Temple University, Philadelphia, PA 19122, USA

**Abstract**
We report inelastic neutron scattering experiments on a single crystal of the intermediate valence compound CePd$_3$. At 300 K the magnetic scattering is quasielastic, with halfwidth $\Gamma$ = 23 meV, and is independent of momentum transfer $Q$. At low temperature, the $Q$-averaged magnetic spectrum is inelastic, exhibiting a broad peak centered near $E_{max}$ = 55 meV. These results, together with the temperature dependence of the susceptibility, 4$f$ occupation number, and specific heat, can be fit by the Kondo/Anderson *impurity* model. The low temperature scattering near $E_{max}$, however, shows significant variations with $Q$, reflecting the coherence of the 4$f$ lattice. The intensity is maximal at (1/2, 1/2,0), intermediate at (1/2,0,0) and (0,0,0), and weak at (1/2,1/2,1/2). We discuss this $Q$-dependence in terms of current ideas about coherence in heavy fermion systems.

The intermediate valence (IV) metals, which are moderately heavy fermion compounds with effective masses $m^*$ in the range 10-50 $m_e$, continue to contribute to our understanding of correlated electron phenomena [1]. The basic physics is that of the Anderson lattice [2], where localized 4$f$ electrons on a periodic array hybridize with a background band of $spd$ electrons, with onsite Coulomb interactions between 4$f$ electrons giving rise to correlated motion. Despite the fact that the 4$f$ electrons sit on a lattice, the Kondo-Anderson *impurity* model (K/AIM) does a good job in describing many of the properties of these materials [3,4]. At low temperatures, however, coherence develops in the 4$f$ lattice giving rise to renormalized band behavior. In YbAl$_3$, a number of anomalies occur below 50 K that have been attributed to the onset of coherence [5].

Because the Kondo physics is dominated by the spin fluctuations, inelastic neutron scattering (INS) is a key experiment for IV metals. Many such measurements have been done on polycrystals, where the spectra are consistent with the K/AIM [6]. Only a small number of single crystals, such as YbInCu$_4$ [7], YbAl$_3$ [8], and CeInSn$_2$ [9] have been studied by INS. In this paper we present data for a single crystal of the IV compound CePd$_3$. We show that the neutron spectra at 300 K are independent of momentum transfer $Q$. This $Q$-independence is an important necessary condition for the applicability of the K/AIM, which provides excellent fits to the susceptibility $\chi(T)$, 4$f$ occupation number $n_f$, and $Q$-averaged magnetic neutron spectrum $\chi''(\Delta E)$ of CePd$_3$. At low temperatures, however, the INS exhibits a $Q$-dependence which requires a theory of the coherent state.

Data for our single crystal of CePd$_3$ are shown in figure 1. The results are very similar to those seen in earlier work on this compound and are very typical of IV metals. The ground state valence $z = 4 - n_f$ has the value 3.75 and the magnetic susceptibility $\chi(T)$ shows a broad maximum near 125 K [10]. We obtain a value $\Gamma \approx 29$ mJ/mol-K$^2$ for the linear coefficient of specific heat [11]. The $Q$-averaged magnetic neutron scattering spectrum peaks at $E_{max}$ = 55 meV [12], implying a Kondo temperature $T_K = E_{max}/k_B$ of order 600 K. The temperature scale $T_{coh}$ 247 $\approx$ 50 K has been identified [13] as that below which coherent renormalized band behavior occurs. The anomalous low temperature upturn in $\chi(T)$ has been shown [10] to be intrinsic and to correlate with the onset of a $5d$ spin susceptibility below 50 K as seen in magnetic form factor measurements[14]; this $5d$ susceptibility is presumably due to coherent $f$-$d$ hybridization.

Earlier INS experiments on CePd$_3$ using a time-of-flight (TOF) spectrometer concluded that most of the spectral weight of the low temperature spin fluctuations resides in a broad peak centered near $E_{max}$ = 53 meV [12]. Measurements on a triple axis spectrometer, however, concluded that the spin fluctuation spectrum could be fit as the sum of two broad peaks centered at $\Delta E$ = 0 and 15 meV [15]. We have speculated [16] that the discrepancy might arise from a significant difference between the zone boundary and zone center scattering.

To clarify the $Q$-dependence of the scattering, we grew an 18 gram single crystal of CePd$_3$ and an 11 gram crystal of LaPd$_3$ by the Czochralski method. The susceptibility, specific heat, and $4f$ occupation number measurements, as well as the K/AIM calculations using the non-crossing approximation (NCA), were carried out in identical fashion to those reported in reference [3]. The samples were aligned with a [1,0,0] direction held vertical. INS experiments were performed on the TOF spectrometers MAPS and MERLIN at the spallation source ISIS at the Rutherford Appleton Laboratory. The incident energy was set at 120 meV for both spectrometers. In the MAPS measurements, the angle between the incident neutron beam and the $x$-axis of the crystal was initially fixed at $\phi = 0$ so that only three components of $\vec{Q}$ - $\Delta E$ space are independent. For example, the Miller index $H = a_0 Q_x/2\pi$ varies with energy transfer $\Delta E$ when the Miller indices $K$ and $L$ are fixed. We also measured on MAPS for $k_i$ along [1,1,0] ($\phi = 45$ degrees); in this situation the decomposition of the horizontal momentum component $Q_H$ onto both of the Miller indices $H$ and $K$ depends on energy transfer.

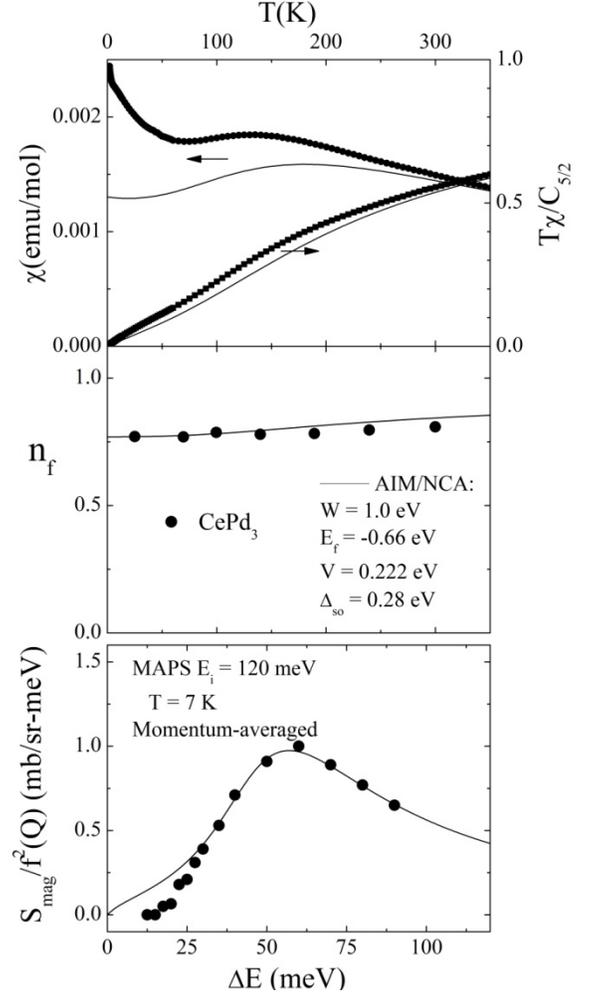

**Figure 1.** The temperature dependence of (a) the magnetic susceptibility $\chi$ and the related effective moment $T \chi /C_{5/2}$, where $C_{5/2}$ is the free ion Curie constant and (b) the $4f$ occupation number $n_f$ of the single crystal of CePd$_3$. (c) The $Q$-averaged magnetic neutron scattering spectrum at $T$ = 7 K. The solid lines are the results of a K/AIM calculation in the NCA, with input parameters given in the plot.

We also measured the scattering of LaPd$_3$ on MAPS to help determine the nonmagnetic background scattering, and we measured both samples at $T$ = 7 and 300 K. In the MERLIN measurements (performed at 5 K) the angle $\phi$ was rotated in discrete steps ("Horace mode") of 2 degrees. Under these circumstances, the scattering intensity over a full 4D volume of $\vec{Q}$ - $\Delta E$ space is determined by interpolation and the intensity can be plotted for any fixed combinations of momentum and energy transfer; for example, constant-$Q$ spectra can be determined. Measurements on both spectrometers were put on an absolute scale by calibrating the scattering against that of a vanadium standard.

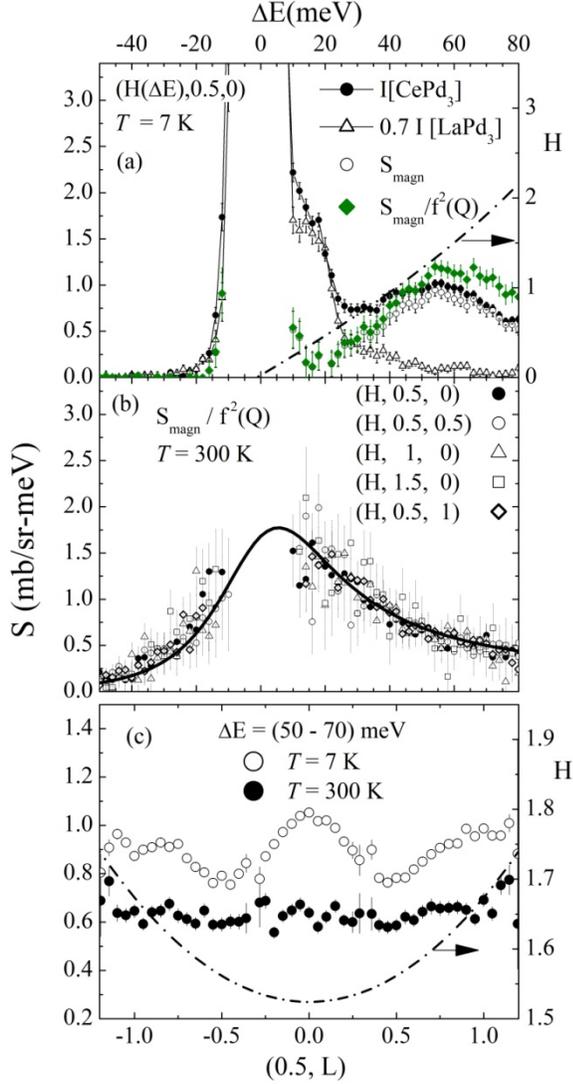

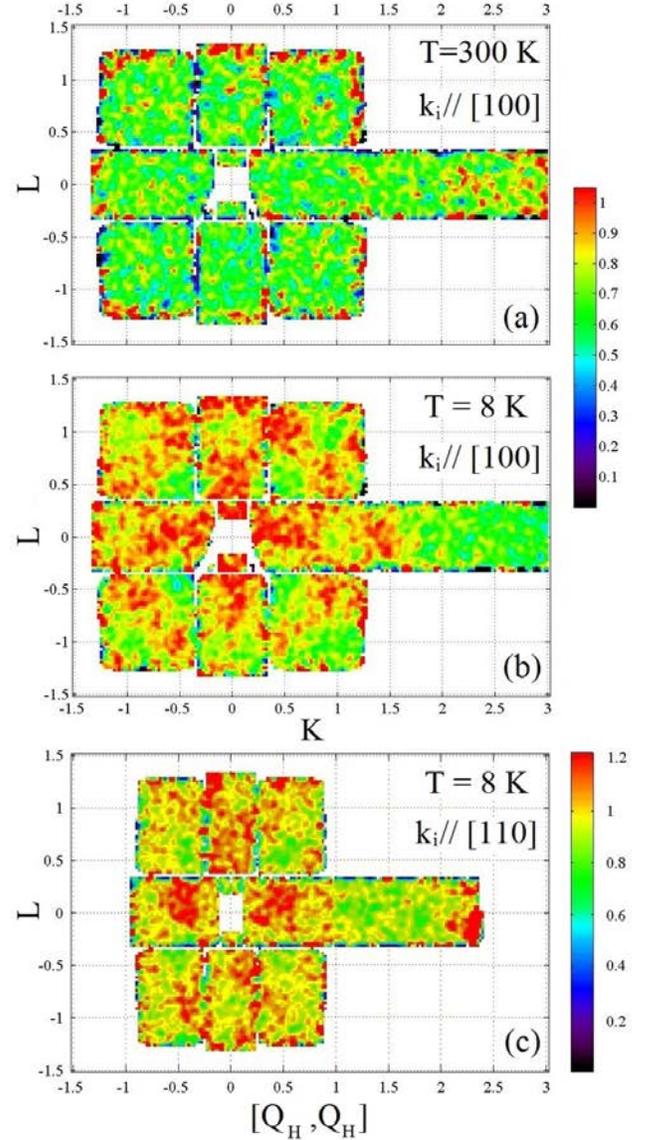

**Figure 2.** (colour online) (a) The inelastic scattering spectra for $CePd_3$ and $LaPd_3$ observed on MAPS at 7K for fixed Miller index $K = 1/2$, $L = 0$. The difference between these two spectra, scaled by the scattering length, represents the magnetic scattering. (b) The room temperature magnetic scattering at a number of fixed values of $K$ and $L$; the scattering is quasielastic (solid line) and independent of momentum transfer $Q$. (c) The scattering versus $L$ for $K = 1/2$ for energy transfer in the range 50-70 meV. The dot-dash lines in (a) and (c) give the variation of the Miller index $H$.

**Figure 3.** (colour online) The scattering intensity (in mb/sr-meV) as measured on MAPS, plotted versus momentum transfer for energy transfer in the range 45-65 meV (a) at 300 K with $k_i$ // [1,0,0], (b) at 7 K with $k_i$ // [1,0,0] and (c) at 7 K with $k_i$ // [1,1,0]. The Miller index $H$ is not fixed in these plots; for example it varies with the magnitude $K^2 + L^2$ in panels (a) and (b). In panels a) and b), $H = 1.50$ for $(K, L) = (0, 1/2)$ and equivalent; i.e. the bright spots in b) correspond to (1/2, 1/2, 0).

The low temperature spectra of $CePd_3$ and $LaPd_3$ at $(K, L) = (1/2, 0)$ as measured on MAPS are compared in figure 2(a). The dashed line shows $H(\Delta E)$. The latter compound exhibits the non-magnetic scattering expected for the former: single phonon scattering below a cutoff of 22 meV [17] and multiple scattering extending from the cutoff to about 40 meV. The magnetic scattering $S_{mag}$ (open circles) in CePd$_3$ is determined by subtracting the data for LaPd$_3$, scaled by a factor of 0.7 to represent the difference in the neutron scattering lengths [18]. The green diamonds represent $S_{mag}/f^{\,2}(Q)$, where the 4$f$ form factor $f^{\,2}(Q)$ has been divided out to give the scattering in the first Brillouin zone. The momentum-averaged scattering of figure 1c was determined by averaging over a number of such spectra for different fixed values of $K$ and $L$. The low temperature scattering peaks near $E_{max}$ = 55 meV; in what follows we will refer to this as the "Kondo scale."

Figure 2(b) exhibits the magnetic scattering at 300 K for a number of fixed values of $K$ and $L$. The solid line in figure 2(b) represents quasielastic (QE) scattering with a halfwidth of 23 meV, similar to that seen earlier in polycrystalline samples [12]. The room temperature scattering is clearly independent of momentum transfer. This can also be seen in figure 3(a), which plots the room temperature scattering intensity for energy transfer in the ($K$, $L$) plane for energy transfer of order $E_{max}$.

On the other hand, the low temperature scattering in the energy range of the Kondo scattering $E_{max}$ shows significant $Q$-dependence. This can be seen in figures 3b and 3c, which plot the scattering at 55 meV versus momentum transfer for $k_i$ // [1,0,0] and [1,1,0]; it can also be seen in the MAPS data of figure 2(c), which plots the scattering versus $L$ at $K$ = 1/2 and $\Delta E$ = 60 ± 10 meV. While the room temperature data is essentially independent of momentum transfer, the low temperature data oscillates with $Q$. In figure 3(b), the scattering is most intense near points equivalent to ($K$, $L$) = (1/2, 0). As mentioned, $H$ varies with $\Delta E$ in figure 2c and varies with $H^{\,2} + K^{\,2}$ in figure 3(b). Bright spots occur at (1.4, 0.5, 0), the weak spots occur at (0.5, 0.5, 0.5) points, and intermediate intensity spots occur at (1.6, 1, 0). In figure 3c, the bright spots are observed at (1.4, -0.6, 0). The proximity of these numbers to integers and half integers suggests that the Kondo scale scattering is strongest at (1/2, 1/2, 0), intermediate at (1/2, 0, 0) and weakest at (1/2, 1/2, 1/2).

To confirm this $Q$-dependence, we have measured the spectra on MERLIN, which allows us to circumvent the non-constancy of $H$ in figures 2 and 3. We plot the low temperature scattering intensity as measured on MERLIN in the (3/2, $K$, $L$) plane as seen near $E_{max}$ (figure 4(a)) and the scattering near $E_{max}$ versus $L$ for ($H$, $K$) = (3/2,0) and (3/2, 1/2) (figure 4(b)). Data for a cut along $L$ with $H$ = 1 and $K$ = 1 (not shown here) indicate that the (0, 0, 0) scattering has a similar intensity to that of the (1/2, 0, 0) scattering. In confirmation of the MAPS data, maxima are seen in the MERLIN plots at all Q that are equivalent to (1/2, 1/2, 0) while minima are seen at (1/2, 1/2, 1/2) points; the intensity at 60 meV for (1/2, 0, 0) and (0, 0, 0) scattering is intermediate.

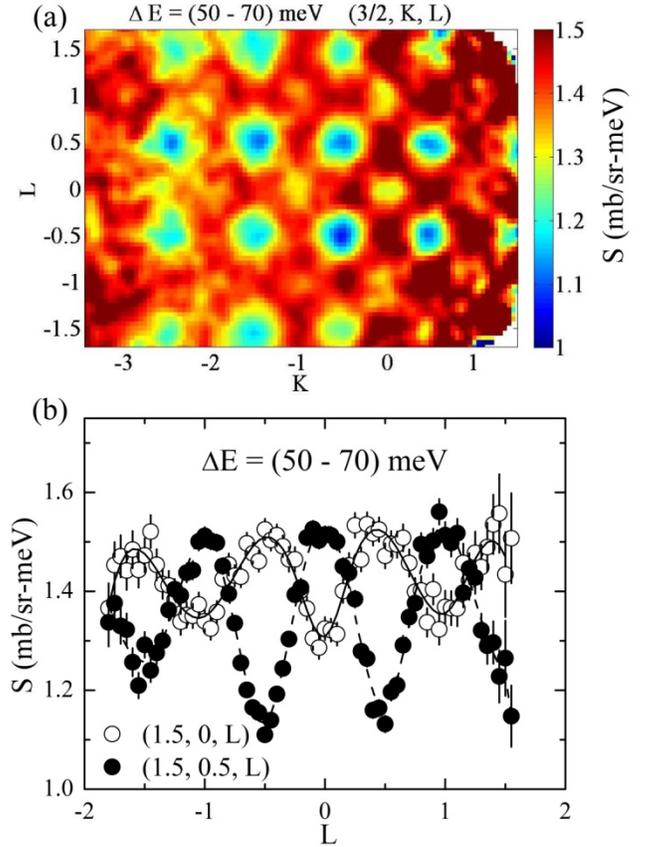

**Figure 4.** (colour online) (a) The low temperature scattering intensity (in mb/sr-meV) versus $K$ and $L$ as measured on MERLIN for $H$ = 3/2 and 50 ≤ $\Delta E$ ≤ 70 meV. (b) Intensity versus $L$ for 50 ≤ $\Delta E$ ≤ 70 meV and for ($H$, $L$) = (3/2, 0) and (3/2, 1/2), as measured on MERLIN. (The lines are guides to the eye.) The scattering near $E_{max}$ is most intense for momentum transfer equivalent to (1/2, 1/2, 0), intermediate for (1/2, 0, 0), and weakest for (1/2, 1/2, 1/2).

Hence we have established that there is significant variation of the Kondo-scale scattering with $Q$ in the low temperature coherent state. On the other hand, the data for the 4f occupation number $n_f(T)$ and the $Q$-averaged magnetic scattering $\chi''(\Delta E)$ can be fit very well by the results of a calculation using the non-crossing approximation (NCA) to the Anderson *impurity* model (figure 1). Apart from the low temperature "tail" (which is a striking coherence effect arising from a correlation between 4f and 5d electrons), the susceptibility is also fit very well by this calculation. The calculation gives 30 mJ/mol-K$^2$ for the specific heat coefficient, which is essentially equal to the measured value 29 mJ/mol-K$^2$. The lack of $Q$ dependence as well as the quasielastic form of the spectrum that we have observed at room temperature (figure 2(b)) are as expected for the K/AIM at elevated temperature. Given that the calculated results depend mainly on two key parameters, the hybridization $V$ and the $f$-level energy $E_f$, and given that the 4f electrons in CePd$_3$ are not impurities but sit on a lattice, the quality of the fits is perhaps surprising, but similar quality fits have been observed for other IV compounds [3,5].

We have argued elsewhere [1] that the applicability of the K/AIM to these periodic compounds is a consequence of the fact that the 4f spin fluctuations are highly localized in space. This is supported by the fact that the $Q$-dependence on the Kondo scale that we do observe at low temperature in CePd$_3$ is not enormous. To show this, we return to the MAPS data, for which we have subtracted the non-magnetic scattering (which we have not done for the MERLIN data), allowing us to compare the magnitude of the scattering at different $Q$. In figure 5, we plot the magnetic spectra for $k_i$ // [1, 0, 0] at fixed values of $(K, L)$ (solid points). The variation of $H$ with energy transfer is shown as the dashed-dot lines. In figure 6, we plot the spectra for $k_i$ // [1, 1, 0]. As mentioned above the decomposition of the momentum transfer $Q_H$ onto $H$ and $K$ depends on energy transfer; a typical example is shown in figure 6(c). In figure 5(d), we plot the magnetic scattering at high symmetry points, determined from the data of Figures. 5 (a)-(c) at the energies where $H(\Delta E)$ takes on an integral or half-integral value and also from the energies in figure 6 where $H$ and $K$ take on integral or half-integral values. This plot shows that the MAPS data agrees well with the MERLIN data in that the Kondo scale scattering (i.e. for $\Delta E \sim 55$ meV) is strongest for (1/2, 1/2, 0), intermediate for (1/2, 0, 0) and (0, 0, 0) and weakest for (1/2, 1/2, 1/2). On the other hand, for energy transfer below 40 meV, the situation reverses, with the (1/2, 1/2, 1/2) scattering being strongest.

We note that the data differ by 20-30 % from the K/AIM fits to the polycrystalline-averaged data (solid lines in figure 5 and 6). Earlier studies along a single zone direction in YbInCu$_4$ [7] and CeInSn$_2$ [9] also concluded that there is only a 20 % increase in intensity between zone center and zone boundary scattering. The shape of the spectra was the same for both $Q$, being essentially of the inelastic Lorentzian form expected based on the K/AIM. In a more recent study of CeInSn$_2$ using polarized neutrons, Murani *et al.* have shown [19] that the $Q$-variation of the 4f form factor varies only by 10 % from that of a 4f local moment. Such a moderate $Q$-dependence is very different from the predictions of the Anderson lattice where the Kondo scale scattering is strongly suppressed at $Q = 0$ (figure 7(a)). On the other hand, the modest $Q$-dependence that we observe is not very different from that expected for a localized ($Q$-independent) spin fluctuation, which helps explain why bulk properties such as $\chi(T)$, $n_f(T)$, $\Gamma$, and $\chi''(\Delta E)$ which depend primarily on the (nearly localized) spin fluctuation continue to follow K/AIM behavior even at low temperature.

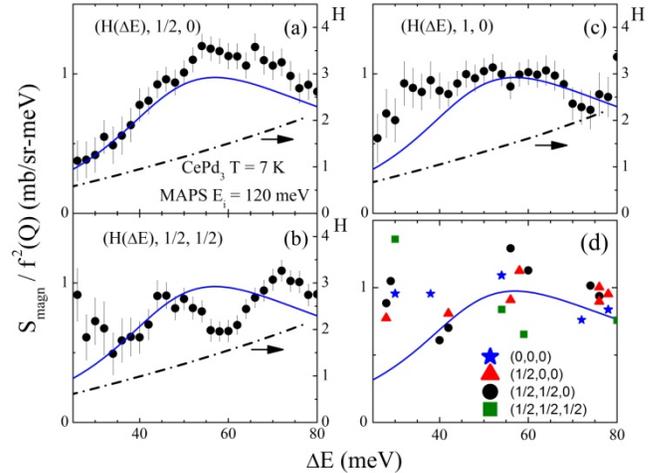

**Figure 5.** The spectra of CePd$_3$ measured on MAPS at $T = 7$ K with sample angle $\phi = 0$ at $(K, L) =$ (a) (1/2, 0), (b) (1, 0), (c) = (1/2, 1/2). The variation of Miller index $H$ with energy transfer is plotted as the dash-dot lines. Data for (a) – (c) at high symmetry points where $H(\Delta E)$ is integral or half integral is plotted in (d). The solid lines are the K/AIM fits to the $Q$-averaged scattering, taken from figure 1 (c).

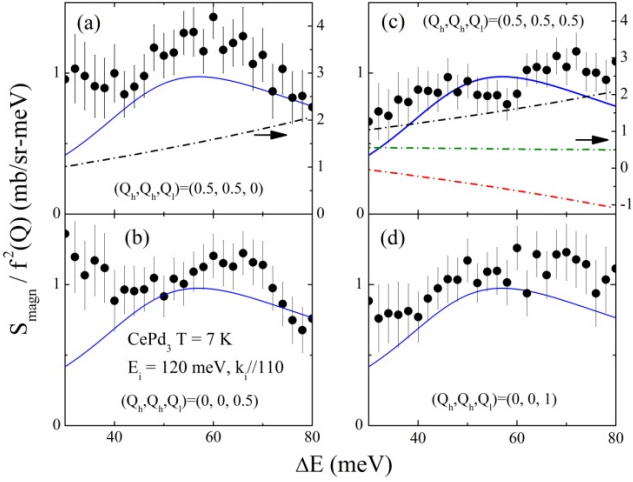

**Figure 6.** The spectra of CePd$_3$ measured on MAPS at $T$ = 7 K at sample angle $\phi$ = 45 degrees at four locations in the detector plane. The Miller indices $H$ and $K$ vary with energy; an example is shown in panel (c). The solid lines are the K/AIM fits to the $Q$-averaged scattering, taken from figure 1(c).

The K/AIM cannot, however, capture properties of the coherent IV ground state that depend critically on the 4$f$ periodicity and coherence. In CePd$_3$ the resistivity behaves as $T^2$ below 50 K [13] and the Hall mobility exhibits an anomaly at the same temperature [20], both of which facts indicate that the coherent Fermi liquid state is stable below a coherence temperature of $T_{coh}$ = 50 K. The low temperature "tail" in the susceptibility develops on the same scale. As mentioned above, this has been shown to be intrinsic behavior, representing correlation between the 5$d$ and the 4$f$ moments -- a truly striking coherence effect. The optical conductivity develops a well-established mid-infrared peak (0.26 eV) at 100 K [21]; this is believed to reflect the onset of the hybridization gap. The full enhancement of the effective mass ($m^*$ = 40 $m_e$), which is determined from the narrow Drude peak in the conductivity and which is indicative of the establishment of the fully renormalized Fermi liquid, appears to occur on the scale $T_{coh}$ [22]. To understand these properties, the low temperature scale on which they occur, and the $Q$-dependence in the magnetic scattering requires a lattice-based theory of the coherent state.

A common argument is that oscillations of the neutron spectra with momentum transfer arise from antiferromagnetic (AF) correlations. This accords with the fact that the scattering is most intense for zone boundary $Q$. In the early work on CeInSn$_2$ [9], this was the explanation given for the oscillation in the scattering intensity. Recently, however, Murani et al. [19] have argued that this is not the case, because the oscillation has a similar magnitude in CeInSn$_2$ where the Kondo temperature is of order $T_K$ ~ 100 K and in CePd$_3$ where $T_K$ ~ 600 K. The scale of antiferromagnetic correlations in cerium compounds is typically less than 10 K, which is much smaller than the 600 K Kondo scale in CePd$_3$, making it unlikely that AF correlations play a significant role. Instead, Murani et al. [19] propose that $Q$-dependent hybridization is responsible for the oscillation.

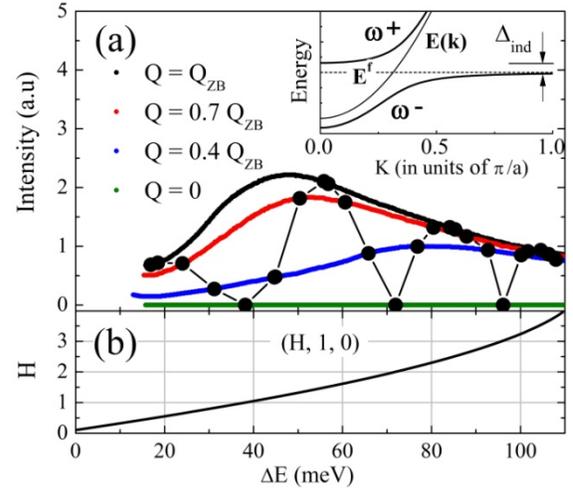

**Figure 7.** (a) Spectra calculated for the Anderson lattice, from Ref [24]. We have scaled the energy axis so that the peak for zone boundary $Q$ corresponds to the value of $E_{max}$ observed experimentally. Inset: Schematic drawing showing the bare bands and the hybridized bands of the Anderson lattice; $\Delta_{ind}$ is the indirect gap. (b) The variation of Miller index $H$ with $\Delta E$ on MAPS for $E_i$ = 120 meV and ($K, L$) = (1, 0). Symbols are drawn on the spectra of a) at the energies where $H(\Delta E)$ takes on the (reduced) value of $Q$ appropriate for each spectrum; the line connecting these points represents the hypothetical spectrum that would be seen on MAPS along the $(H(\Delta E), K, L)$ direction based on the theory.

As mentioned in the introduction, the most basic lattice-based theory of IV compounds is the Anderson Lattice Model (ALM). As shown in figure 7(a) (inset), this predicts an indirect gap $E_{ind}$, which has the magnitude of the Kondo energy, between zone center and zone boundary due to the hybridization of a narrow *f* level with a broad conduction band. The inelastic neutron scattering arises from particle-hole excitations, and is most intense for transitions between regions of high density of states in both the occupied and the unoccupied bands. For the Anderson lattice, this occurs for $\Delta E \sim E_{ind}$ and for zone boundary momentum transfer $Q_{ZB}$. Such behavior is seen in Kondo insulators such as $YbB_{12}$ [23]. The calculated scattering decreases rapidly in intensity and shifts to higher energy transfer as $Q$ decreases; indeed, for $Q = 0$ it becomes negligible on the scale of the indirect gap and is large on the scale of the direct gap. This is shown in figure 7(a), where we plot the calculations of Aligia and Alascio [24] for the Anderson lattice.

To compare the ALM calculations to our data, we have scaled the energy axis of the calculation so that the zone boundary scattering is on the same energy scale as $T_{max}$ for $CePd_3$. The MAPS data for the cut $(H(\Delta E), 1, 0)$ (figure 5(c)) traverses the zone center and zone boundary along the [1,0,0] direction. To make a rough comparison to this data, we have drawn symbols on the curves of figure 7a at the energies where $H(\Delta E)$ takes on the reduced value of $Q$ appropriate for each spectrum. The resulting curve (thin line through the symbols) is very different from the data of figure 5(c). In particular, the oscillation with $H(\Delta E)$ is much smaller experimentally than predicted by the theory, essentially because the scattering on the Kondo scale does not vanish at zone center points. (The oscillation seen in figure 5(b) for the $(H, 1/2, 1/2)$ cut represents a variation which occurs entirely on the zone face between (1/2, 1/2, 0) and (1/2, 1/2, 0); it occurs because the Kondo scale scattering at (1/2, 1/2, 1/2) is smaller than at the other high symmetry points.)

Hence there are serious difficulties with the standard Anderson lattice prediction for the dynamic susceptibility. Perhaps the most obvious problem arises from the fact that the calculation of figure 7a is carried out for a single parabolic conduction band, crossed by a single mono-energetic *f* level. This gives rise to the hybridized bands of figure 7(a) inset, with a single hybridization gap. In a real IV material, the energy gaps are governed by the bandstructure, and correlated band theory is needed to capture the behavior. The hybridized *f* bands can be much flatter over broad intervals of *k*-space than in the simple Anderson lattice, and indirect and direct gaps can occur between different high symmetry points. This can be seen in the recent band calculation of Sakai [25] for $CePd_3$ which uses dynamic mean field theory (DMFT) to include correlations on top of the standard LDA treatment (figure 8). Unlike the Anderson lattice, there is a direct gap of about 26 meV between occupied and unoccupied states at the *Γ* point, allowing $Q = 0$ transitions. There are also indirect gaps of order 30 and 36 meV between *Γ* and *M* and between *R* and *X* corresponding to a (1/2, 1/2, 0) transition. The *R-Γ* and *Γ-R* gaps corresponding to a (1/2, 1/2, 1/2) transition are at 20 and 85 meV respectively.

Hence, while the band calculation predicts somewhat smaller energies for the excitations than is seen in our data, it predicts the correct order of magnitude for the Kondo energy scale, it shows that the gaps can vary with momentum transfer, it shows that $Q = 0$ scattering can be significant rather than vanishingly small as predicted by the ALM, and it demonstrates that intense Kondo-scale scattering should be seen at (1/2, 1/2, 0). It also suggests that the (1/2, 1/2, 1/2) scattering will occur at energies that are either substantially larger or smaller than the scattering at (1/2, 1/2, 0). Indeed, it is plausible that the enhanced (1/2, 1/2, 1/2) scattering seen at 30 meV in figure 5(d) represents the lower energy *R-Γ* scattering.

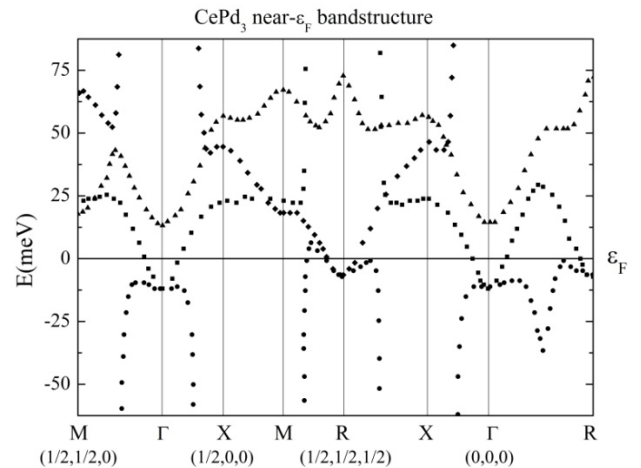

**Figure 8.** The renormalized band structure of $CePd_3$ calculated in LDA+DMFT. Adapted from Sakai [25].

To improve the comparison to band theory will require two advances. The first is that the scattering should be determined with controlled momentum transfer, as in our MERLIN experiment, but over a broader range of energy transfer than shown here. To determine the magnetic scattering, the nonmagnetic scattering should be determined and subtracted from the measured spectra. Second, a calculation of the $Q$- and $E$-dependent dynamic susceptibility $\chi''(Q,E)$ should be performed, based on the correlated band theory. Such a calculation would include not only the particle-hole excitations between the centroids of the bandstates, but the full inelasticity of the momentum resolved spectral functions. The point is that the bandstates $\varepsilon(k)$ of the renormalized band theory are only sharp in energy at low temperature and for energies close to the Fermi level. As the temperature increases, or as the distance of the energy from the Fermi level increases, the spectral functions rapidly become quite broad in energy. This broadening results from the strong inelastic scattering that is inherent in correlated electron systems. For our purposes, it has two major consequences. The first is that as the temperature is raised, the crossover to incoherent scattering occurs at a relatively low temperature. An indication of this is that we see $Q$-independent quasielastic scattering spectra already at 300 K, which is $\sim T_K/2$. The second consequence is that the low temperature spectra at finite $\Delta E$ will be considerably broadened in both $Q$ and $E$ compared to the spectra calculated without including the inelasticity. These two features have been seen clearly in calculations for the Anderson lattice [26] and for the Hubbard model [27], and in LDA+DMFT calculations for real materials [28]. We believe that it is the broadening in $Q$ and $E$ that causes the inelastic neutron scattering spectra that are observed in IV systems to closely resemble those expected for the Anderson *impurity* model.

As a final point, we note that figure 1 shows that, apart from the low temperature 5$d$ contribution, the Kondo-scale scattering accounts for most of the d.c. susceptibility. Hence our data supports the contention of Murani *et al.* [12] that the primary spectral weight of spin fluctuations is on the scale $E_{max}$. Our earlier conjecture [16] that the zone center scattering is shifted to lower energy is not borne out by the current analysis; on the Kondo scale, the zone center data is not very different from the zone boundary data. Our procedure for subtracting the nonmagnetic scattering (Figure 2 (a)) is not adequate, however, to determine whether magnetic excitations exist at energies below 30 meV where the nonmagnetic scattering is important. Hence we cannot rule out the contention of Shapiro *et al.* [15] that significant scattering occurs at lower energy. Indeed, a study using polarized neutrons and a polycrystalline sample of $CePd_3$ suggested there might be a 15 meV excitation in addition to the one at 55 meV [29]. We believe that the whole issue of the low energy scattering is very much an open question for IV compounds. Most existing studies of single crystals of these compounds [7, 8, 9, 19] have explored the scattering on the Kondo scale, where the scattering corresponds to *inter*band scattering between the renormalized bands above and below the Fermi energy. In addition, *intra*band scattering for particle-hole excitations in occupied bands that cross the Fermi level should be observed. Such excitations give rise to dispersive Fermi liquid scattering with $\varepsilon \sim Q\, v_F$ where $v_F$ is renormalized by the excitations. While such scattering has been observed in the heavy fermion compound $UPt_3$ [30], to the best of our knowledge it has not been seen in any IV compound to date. The study of the excitations at energies below the phonon cutoff would be best addressed using polarized neutrons and a single crystal sample.

In conclusion, we have shown that much of the behavior of the IV compound $CePd_3$ can be understood as that of a Kondo/Anderson *impurity*. At low temperatures, however, the Kondo scale magnetic scattering exhibits a momentum dependence that can be understood as arising from coherent 4$f$ bands. A full comparison of the Kondo-scale scattering to the predictions of correlated band theory, and a determination of the momentum- and energy-dependence of the scattering below the phonon cutoff (20 meV), remain open questions that will require future measurements to resolve.

Work at the University of California, Irvine (UCI), Temple University, Argonne National Laboratory (ANL), Lawrence Berkeley National Laboratory (LBNL), and Los Alamos National Laboratory (LANL) was supported by the Office of Basic Energy Sciences (BES) of the U.S. Department of Energy (DOE). The work was funded by awards DE-FG02-03ER46036 (UCI), DE-FG02-84ER45872 (Temple), DE-AC02-05CH11231 (LBNL), and DE-AC02-06CH11357 (ANL). Research at Oak Ridge National Laboratory (ORNL) and at the Stanford Synchrotron Radiation Lightsource (SSRL) was sponsored by the Scientific User Facilities Division of BES/DOE.